\documentclass[twocolumn,showpacs,groupedaddress]{revtex4}
\usepackage{graphicx}

\begin{document}

\title{Geometric phases in a scattering process}
\author{H. D. Liu and  X. X. Yi}
\affiliation{School of Physics and Optoelectronic Technology, Dalian
University of Technology, Dalian 116024, China}

\date{\today}

\begin{abstract}
The study of geometric phase in quantum mechanics has so far be
confined to discrete (or continuous) spectra and trace preserving
evolutions. Consider only the transmission channel, a scattering
process with internal degrees of freedom is neither a discrete
spectrum problem nor a trace preserving process. We explore the
geometric phase in a scattering process taking only the transmission
process into account. We find that the geometric phase can be
calculated by the some method as in an unitary evolution. The
interference visibility depends on the transmission amplitude. The
dependence of the geometric phase on the barrier strength and the
spin-spin coupling constant is also presented and discussed.
\end{abstract}

\pacs{03.65.Bz, 11.15.-q}\maketitle

Berry's phase was originally introduced for bound states that an
(discrete) eigenstate of the Hamiltonian would accumulate a
geometric phase\cite{berry84}, when the evolution of the system is
adiabatic. This Berry's phase  provides us a very deep insight on
the geometric structure of quantum mechanics and gives rise to
various observable effects. The concept of the Berry phase has now
become a central unifying concept in quantum mechanics, with
applications in fields ranging from chemistry to condensed matter
physics \cite{shapere89}. Recently the concept of Berry phase  has
been renewed  and generalized for mixed
states\cite{erik00,tong04,yi04}. All these studies  have been
confined to discrete spectra.

For continuous spectrum, there are two things that can distinguish
the geometric phase from  bound states. (1) We always have
non-Abelian gauge as a connection due to the degeneracy in this
situation \cite{newton94}; (2) The distortion of the Hamiltonian can
not limited to a finite set of parameters, and hence we have to take
into account the problem in an infinite-dimensional space. With
these observations, the geometric phase factor  has been considered
for continuous spectra in \cite{newton94}, showing that the factor
is exactly the scattering matrix. In Ref. \cite{ghosh96}, the
scattering phase shift is defined in a way analogous to the
adiabatic phase for bound states. This method works when reflection
is negligible. By defining a virtual gap for the continuous spectrum
through the notion of eigen-differential and using the differential
projector operator, an explicit formula for a generalized
geometrical phase is derived in terms of the eigenstates of the
slowly time-dependent Hamiltonian\cite{maamache08}. These studies,
in contrast with the case of discrete spectra, are all for systems
with continuous spectra.

\begin{figure}
\includegraphics*[width=0.7\columnwidth,
height=0.5\columnwidth]{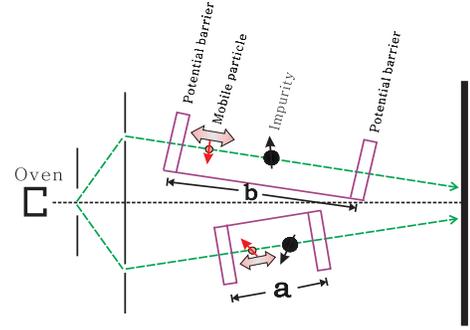}
 \caption{(Color online) Illustration
of a gedanken setup. A mobile particle can propagate along a wire in
each path. A quantum impurity and two narrow potential barriers lie
at $-\frac a 2$ and $\frac a 2$ in one path, and at $-\frac b 2$ and
$\frac b 2$  in another. Once the mobile particle injected into one
of the path, it undergoes multiple reflections between the barriers
and impurity. Eventually, the mobile particle transmitted froward or
reflected back. Consider only the transmission channel, this
scattering process is not of trace-preserving.  $a$ ($b$) is the
distance between the two barriers that we will refer to the width of
 structure in the text. }\label{f1}
\end{figure}

A scattering process with particles that have (pseudo) spin degrees
of freedom is a typical phenomenon  different from the
aforementioned: The (discrete) internal spin degrees of freedom of
the scattering particles inevitably couple to the (continuous)
motional dynamics \cite{tang95}. Hence such processes affect the
state of the colliding spins according to quantum maps, instead of
unitary operations. This makes the geometric phase acquired in such
scattering processes distinct and  interesting. Our main motivation
in the present paper is to study the geometric phase in a scattering
process with pseudo spin degrees of freedom. To tackle the problem,
we focus on a gedanken setup consisting a quantum impurity, a mobile
particle and two narrow potential barriers in each path of the
double-slit, as shown in Fig. \ref{f1}.

The mobile spin-1/2 particle $e$ can propagate along the  1D path. A
quantum impurity $I$, modeled as a spin-$S$ scatterer, lies at
$x=0$, whereas two narrow potential barriers are located at $x=\pm
x_0$ (the $x$-axis is along the path, $x_0=a/2,b/2$ in Fig.\ref{f1}
for the two paths, respectively). The Hamiltonian for each path
reads \cite{helman85,bose10} (we set $\hbar\!=\!1$ throughout)
\begin{equation}
\label{H} H=\frac{p^{2}}{2m} + J\delta(x) \vec{s}\cdot\vec{S}+ G
\left [ \delta(x-x_0)+\delta(x+x_0) \right ],
\end{equation}
where $m$ and $p$ are the effective mass and momentum operator of
$e$, respectively, $\vec{s}$ and $\vec{S}$ stand respectively for
the spin operators of $e$ and $I$,   $J$ is a spin-spin coupling
constant and $G$ is the potential-barrier strength. The above
paradigmatic model naturally matches within a solid-state scenarios
such as a 1D quantum wire \cite{davies98} or single-wall carbon
nanotube \cite{devoret97} with an embedded magnetic impurity or
quantum dot\cite{ciccarello07}. Potential barriers are routinely
implemented through applied gate voltages or heterojunctions.

Clearly, all of the scattering probability amplitudes are spin
dependent due to the spin-spin contact potential $J\delta(x)
\vec{s}\cdot\vec{S}$ in the Hamiltonian. As the overall spin space
is $D$-dimensional ($D=[2\times(2S+1)]$), the effect of scattering
is fully described by two $D\times D$ matrices whose generic
elements respectively represent the amplitudes of reflection and
transmission. These matrices can be derived by noting that the
squared total spin of $e$ and $I$ as well as its projection along
the $z$-axis are conserved. This entails that the dynamics within
the singlet and triplet subspaces are decoupled. Consider only the
transmission channel and assume that the injected   state is
\begin{equation}
|\varphi_{in}\rangle=e^{ikx}|\uparrow\rangle\otimes|\phi_m\rangle,
\end{equation}
the transmitted  state takes,
\begin{eqnarray}
|\varphi_{out}\rangle
&=&e^{ikx}t_\uparrow|\uparrow\rangle\otimes|\phi_m\rangle
+e^{ikx}t_\downarrow |\downarrow\rangle\otimes|\phi_{m+1}\rangle,
\end{eqnarray}
where $t_{\uparrow}$ and $t_{\downarrow}$  are the probability
amplitudes for transmission with spin up and down, respectively.
$|\phi_m\rangle$ are the eigenstates of $S_z$ (the $z$-component of
$\vec{S}$), i.e., $S_z|\phi_m\rangle=m|\phi_m\rangle,$ and
$k=\sqrt{2mE}$ with $E>0$ being  the energy  of the injected
particle. $|\uparrow\rangle$ and $|\downarrow\rangle$ denote the
eigenstates of $s_z$ for the mobile particle. The dependence of
$t_{\uparrow}$ and $t_{\downarrow}$ on $G$, $J$ and $x_0$ can be
established by
\begin{equation}
t_{\uparrow}=t_{\uparrow}(x_0)=\{1+i[\chi-(m+1)j']\}/\Delta,
\end{equation}
\begin{equation}
t_{\downarrow}=t_{\downarrow}(x_0)=-ij'F/\Delta,
\end{equation}
where $j'=|W|^2j/(2\kappa)$,
$\Delta=(1+i\chi)[1+i(\chi-j')]+S(S+1)j'^2$ and
$F=[(S-m)(S+m+1)]^{1/2}$, with $
W=1+g\sin(2\kappa\alpha)+i2g\sin^2(\kappa\alpha), $ and $
\chi=2g[g\sin(2\kappa\alpha)+\cos(2\kappa\alpha)]. $ To simplify the
problem, the following dimensionless quantities were defined:
$j=J/(2a_B\epsilon)$, $\kappa=ka_B=\sqrt{E/\varepsilon}$,
$g=G/(2ka_B\varepsilon)$, and $\alpha=x_0/a_B.$ Here $a_B$ is the
Bohr radius,  $\varepsilon=1/(2m) a_B^2$ and $2x_0$ is the distance
between the two potential barriers, which we will call the width of
structure in this paper.

Consider a situation where the width of the structure  on each path
is different but the spin-spin coupling constant and the barrier
strength on both paths are the same. We have interests in the phase
difference between the mobile particles transmitted through
different paths. This phase difference consists of a dynamical phase
and a geometrical part. Our task here is to extract the geometric
phase from the total part $\Gamma=\arg\langle
\varphi_{out}(a)|\varphi_{out}(b)\rangle.$ This can be done by
either parallel  transport of the state  or canceling the dynamical
phase. The parallel transport condition in this case is $\Im
\langle\varphi_{out}(x_0)|\frac{\partial}{\partial
x_0}|\varphi_{out}(x_0)\rangle=0$, leading to the geometric phase in
the scattering process,
\begin{widetext}
\begin{equation}
\gamma_s=\arg\left(\langle\varphi_{out}(a)|\varphi_{out}(b)\rangle
e^{-i\Im
(\int_a^b\frac{\langle\varphi_{out}(x_0)|\frac{\partial}{\partial
x_0}|\varphi_{out}(x_0)\rangle
}{\langle\varphi_{out}(x_0)|\varphi_{out}(x_0)}dx_0)}\right),
\label{gps}
\end{equation}
\end{widetext}
where $\Im(...)$ denotes the imaginary part of $(...).$ We now prove
that $\gamma_s$ defined in Eq. (\ref{gps}) is geometric, i.e., it
only depends on the trajectory traced out by
$|\varphi_{out}(x_0)\rangle$. Define a quantum map by
\begin{equation}
M(b,a)=|\varphi_{out}(b)\rangle\langle \varphi_{out}(a)|,
\end{equation}
the total phase  $\Gamma$ acquired in the scattering process can be
written as $\Gamma=\arg\langle
\varphi_{out}(a)|M(b,a)|\varphi_{out}(a)\rangle.$ Notice that
\begin{equation}
\bar{M}(b,a)=M(b,a)e^{i\beta(b,a)}|\varphi_{out}(a)\rangle\langle
\varphi_{out}(a)| \label{MbM}
\end{equation}
with real parameters $\beta(b,a)$ and $\beta(a,a)=0$ gives  the same
state $|\bar{\varphi}_{out}(b)\rangle$, since
$|\bar{\varphi}_{out}(b)\rangle=e^{i\beta(b,a)}|\varphi_{out}(b)\rangle$
differs from $|\varphi_{out}(b)\rangle$ only in  an overall phase
$\beta(b,a)$. Parallel transport condition  $\Im
\langle\varphi_{out}(x_0)|\frac{\partial}{\partial
x_0}|\varphi_{out}(x_0)\rangle=0$ leads to
\begin{equation}
\Im \langle
\varphi_{out}(0)|\bar{M}^{\dagger}(x_0,0)\frac{\partial}{\partial
x_0}\bar{M}(x_0,0)|\varphi_{out}(0)\rangle=0. \label{ptc}
\end{equation}
Substituting Eq. (\ref{MbM}) into Eq.(\ref{ptc}), we have
\begin{equation}
\beta(b,a)=-\int_a^b\frac{\Im \langle
\varphi_{out}(x_0)|\frac{\partial}{\partial
x_0}|\varphi_{out}(x_0)\rangle}{\langle\varphi_{out}(x_0)|\varphi_{out}(x_0)\rangle}
dx_0.
\end{equation}
This completes the proof. For our scattering problem,  simple
algebra yields,
\begin{widetext}
\begin{eqnarray}
\gamma_s=\arg\left [ ( t_{\uparrow}^*(a)
t_{\uparrow}(b)+t_{\downarrow}^*(a)
t_{\downarrow}(b))e^{-i\int_a^b\frac{1}{|t_{\uparrow}|^2+|t_{\downarrow}|^2}(|t_{\uparrow}|^2\frac{\partial
\phi_{\uparrow}}{\partial x_0}+|t_{\downarrow}|^2\frac{\partial
\phi_{\downarrow}}{\partial x_0})dx_0}\right ].\label{gpsd}
\end{eqnarray}
\end{widetext}
Here, $\phi_{\uparrow(\downarrow)}$ was defined by
$$\tan\phi_{\uparrow(\downarrow)}\equiv\frac{t^I_{\uparrow(\downarrow)}}{t^R_{\uparrow(\downarrow)}}.$$
$t_{\uparrow (\downarrow)}^I$ and $t_{\uparrow (\downarrow)}^R$
denote the imaginary and real part of  $t_{\uparrow (\downarrow)},$
respectively. The geometric phase given in Eq.(\ref{gpsd})
represents  the difference in geometric phase for the mobile
particle transmitted through the two paths. We will show later that
it coincides with the geometric phase acquired in an unitary
evolution treating the width as time $t$.

\begin{figure}
\includegraphics*[width=0.7\columnwidth,
height=0.5\columnwidth]{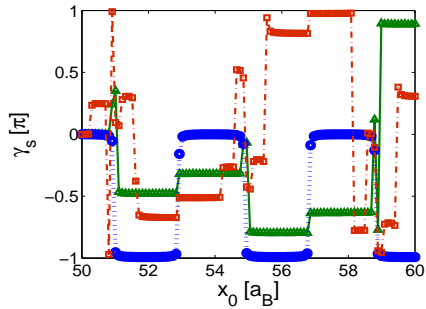}
 \caption{(Color online) The geometric phase $\gamma_s$ versus the
width differences. Parameters chosen are: $J\rightarrow 0$
 $(10^{-6})$ for blue circle; $J=11$ for red square and $J=50$ for
green triangle. The other parameters: $k=0.8$, $a_B=1$, $G=10$,
$m=-\frac 1 2,$ $\varepsilon=1$, $S=\frac 1 2.$}\label{f2}
\end{figure}

\begin{figure}
\includegraphics*[width=0.7\columnwidth,
height=0.5\columnwidth]{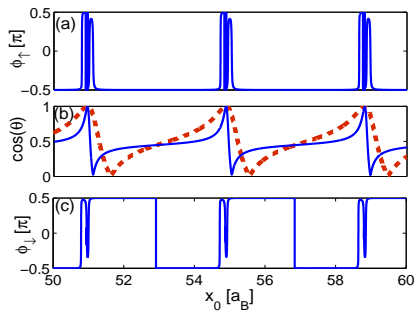}
 \caption{(Color online) Angle $\varphi_{\uparrow}$ and
 $\varphi_{\downarrow}$, and $\cos(\theta)$ as a function of the
 width differences.
 $J=50$ was taken for the plot. The red dashed line in (b) is for  $J=11$.
The other parameters chosen are  the same as in Fig.\ref{f2}.
$\cos\theta$ was defined by $\cos
\theta\equiv\frac{|t_\uparrow|}{\sqrt{|t_\uparrow|^2
+|t_\downarrow|^2}}$.}\label{f3}
\end{figure}
\begin{figure}
\includegraphics*[width=0.7\columnwidth,
height=0.5\columnwidth]{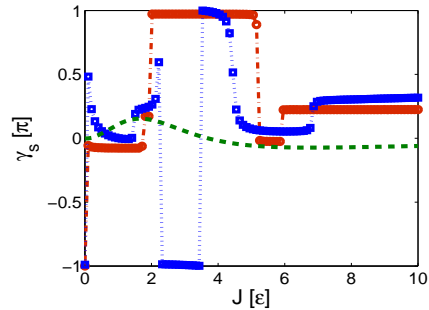} \caption{(Color online)
$\gamma_s$ versus spin-spin coupling constant $J$.  $G=0.1$ for
green dashed line,  $ G=9$ for blue square line, and  $ G=30 $ for
red circle line.  For other parameters, see Fig. \ref{f2}. The width
difference between the two pathes is $60a_B$. }\label{f4}
\end{figure}
\begin{figure}
\includegraphics*[width=0.7\columnwidth,
height=0.5\columnwidth]{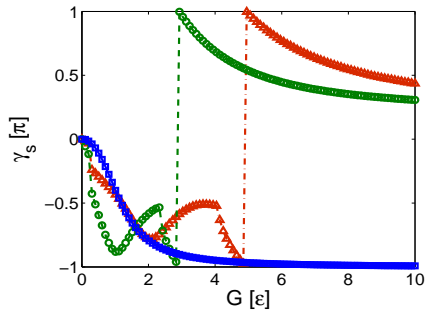}
 \caption{(Color online) $\gamma_s$ as a function of the
 barrier strength. $J\rightarrow 0$ for blue square line, $J=11$
 for green circle, and  $J=30$  for red triangle. The width
difference between the two pathes is $60a_B$.
 }\label{f5}
\end{figure}

We have performed numerical calculations for Eq.(\ref{gpsd}),
results are presented in Fig.\ref{f2}-- Fig.\ref{f5}. For
simplicity, $S=\frac 1 2 $ was specified without loss of generality.
Fig.\ref{f2} shows the dependence of the geometric phase $\gamma_s$
on the width difference (i.e., $b-a$ in Fig.\ref{f1}) on the two
paths for different spin-spin coupling constant. We find that the
mobile particle acquires either $0$ or $-\pi$ geometric phase when
$J\rightarrow 0$ ($J=10^{-6}$ was taken for the plot). Sharp changes
in the geometric phase happen periodically, regardless of what value
$J$ takes. Moreover we find that the geometric phase change its
value only at the points where $t_{\uparrow}$ and $t_{\downarrow}$
change abruptly, as shown in Fig.\ref{f3}. We observe three
resonances from Fig.\ref{f3}, corresponding to $\cos\theta =1.$ As
the spin-spin coupling constant $J$ approaches the barrier strength
$G$, the resonance region becomes wide (see the red-dashed line in
Fig.\ref{f3}(b)).
Further examination shows that these points coincide with the
condition for resonant energies given by $\cot(2\kappa\alpha)=-g$
(i.e., $|t_{\uparrow}|=1$). The spin-spin coupling smooth the
sharpness of the changes, this is due to the broadening of the
energy resonance (see Fig. \ref{f3}, red dashed line). The
dependence of the geometric phase on the spin-spin coupling is shown
in Fig.\ref{f4}. Note that $\gamma_s=0$ when $G=0$,  which  is not
shown on the figure. This can be easily interpreted in the limit of
$g\rightarrow 0$. In this limit, $t_{\uparrow}\simeq (1-i
0.5j^{\prime})/(1-ij^{\prime}+1.5j^{\prime 2}),$
$t_{\downarrow}\simeq (-i j^{\prime})/(1-ij^{\prime}+1.5j^{\prime
2}).$ Clearly, both $t_{\uparrow}$ and $t_{\downarrow}$ do not
depend on the width of the structure, thus the system can not
acquire a geometric phase with $G=0$. This is, however, not the case
for $J=0$ as Fig.\ref{f5} shows. In limit of $J=0$,
$t_{\uparrow}\simeq
 \frac{1}{1+i\chi}$, and $t_{\downarrow} \simeq 0.$ As $\chi$ depends
 on the width, the geometric phase in this case is,
\begin{equation}
\gamma_s=\arg\left ( t^*_{\uparrow}(a) t_{\uparrow}(b)
e^{-i(\phi_{\uparrow}(a)-\phi_{\uparrow}(b))} \right ).
\end{equation}
In the strong spin-spin coupling ($J\rightarrow \infty$) and large
barrier strength limit ($g\rightarrow \infty$), we have
$\phi_{\uparrow}=\phi_{\downarrow}$ and
$t_{\downarrow}=2t_{\uparrow},$ this leads to the geometric phase,
\begin{equation}
\gamma_s=\arg\left ( t^*_{\uparrow}(a) t_{\uparrow}(b)
e^{-i2(\phi_{\uparrow}(a)-\phi_{\uparrow}(b))} \right ).
\end{equation}

Now we are in a position to explore what is the difference between
the normalized and non-normalized transmitted state, in terms of
geometric phase. To this end, we define
$$
\cos\theta\equiv\frac{|t_\uparrow|}{\sqrt{|t_\uparrow|^2
+|t_\downarrow|^2}},
$$
the transmitted state can be rewritten as,
\begin{equation}
|\varphi_{out}\rangle=\cos\theta
e^{i\phi_\uparrow}|\uparrow\rangle\otimes|\phi_m\rangle+\sin\theta
e^{i\phi_\downarrow}|\downarrow\rangle\otimes|\phi_{m+1}\rangle.
\end{equation}
We point out  that by the conservation of  current probability,
$|t_{\uparrow}|^2+|t_{\downarrow}|^2=1-|r_{\uparrow}|^2-|r_{\downarrow}|^2\leq
1.$ Here we  consider only the transmission channel, and the
transmitted state has been normalized, this would  only affect the
visibility of the interference fringes  but not  shift the patterns.
By the definition of geometric phase for an  unitary evolution, we
have
\begin{widetext}
\begin{equation}
\gamma^{\prime}_s=\arg\left((\cos\theta(b)\cos\theta(a)e^{i(\phi_{\uparrow}(b)-\phi_{\uparrow}(a))}+
\sin\theta(b)\sin\theta(a)e^{i(\phi_{\downarrow}(b)-\phi_{\downarrow}(a))})
e^{-i\int_a^b(\dot{\phi}_{\uparrow}\cos^2\theta+\dot{\phi}_{\downarrow}
\sin^2\theta)dx_0}\right),
\end{equation}
\end{widetext}
where $
\dot{\phi}_{\uparrow}\equiv\frac{\partial\phi_{\uparrow}}{\partial
x_0},~\dot{\phi}_{\downarrow}\equiv\frac{\partial\phi_{\downarrow}}{\partial
x_0}. $  Recall that the real part of
$\langle\varphi_{out}(a)|\varphi_{out}(b)\rangle$ represents the
visibility of the interference pattern, we conclude that the
geometric phase for the non-normalized and normalized transmitted
state are the same, namely, $\gamma^{\prime}_s=\gamma_s$. One may
concern about the observation of the geometric phase, inparticulare
worry about the separation of the geometric phase from the total
phase. In general, by varying the width difference $(b-a)$, it is
possible to make the dynamics part of phase the same for the two
beams.

In conclusion, the geometric phase in a scattering process is
studied in this paper. Consider only the transmission channel, the
scattering process is neither a trace-preserving dynamics nor a
discrete spectrum problem. Instead it concerns the coupling between
the internal degrees of freedom and the motional dynamics, and it
can be described by quantum map to replace the unitary evolution. We
have defined and calculated the geometric phase in such a process
and show the dependence of the geometric phase on the spin-spin
coupling constant and the barrier strengths. Possible observation of
the geometric phase is suggested and discussed.
\ \ \ \\
This work is supported by NSF of China under grant Nos 61078011 and
10935010.

\end{document}